# Nonextensivity in nonequilibrium plasma systems with Coulombian long-range interactions

Du Jiulin

*Department of Physics, School of Science, Tianjin University, Tianjin 300072, China*

E-mail: jldu@tju.edu.cn

**Abstract** The nonextensivity in a non-isothermal plasma system with the Coulombian long-range interactions is studied in the framework of Tsallis statistics. We present for first time a mathematical expression of the nonextensive parameter $q$ based on the mathematical theory about the generalized Boltzmann equation and the $q$-H theorem and the Maxwellian $q$-velocity distribution. We obtain a new physical explanation for $q{\neq}1$ concerning the nature of non-isothermal configurations in plasma systems with Coulombian long-range interactions. We also provide one illustration for Almeida's theorem (*Physica A* **300**(2001)424) from the kinetic analyses of plasma systems, which means that Tsallis statistics might be a suitable statistics for the description of a nonequilibrium system with a temperature gradient in it.

**PACS** number(s): 05.70.Ln, 51.10.+y, 05.20.-y, 05.90.+m

Almost all the systems treated in statistical mechanics with the Boltzmann-Gibbs (B-G) statistics have usually been extensive, this property holds for systems with short-range interparticle forces. When we deal with systems with long-rang interparticle forces such as Newtonian gravitational forces and Coulomb electric forces, where nonextensivity holds, the B-G statistics may need to be generalized for their statistical description. In addition to the long-range interacting systems, all dissipative systems, which have nonvanishing thermodynamic currents, appear not to be extensive [1]. In recent years, a nonextensive generalization of B-G statistical mechanics known as "Tsallis statistics" has focused significant attention [2-5]. This generalization was by constructing a new expression of entropy, $S_q$, that depends on the nonextensive parameter $q$ different from unity [6] by the form

$$S_q = \frac{k}{1-q}(\sum_i p_i^q - 1) \qquad (1)$$

where $k$ is the Boltzmann constant, $p_i$ the probability that the system under consideration is in its $i$th configuration such that $\sum_i p_i = 1$, and $q$ a positive parameter whose deviation from unity is considered for describing the degree of nonextensivity of the system. The celebrated Boltzmann entropy $S_B$ is recovered perfectly from $S_q$ if $q$ =1. Thus, Tsallis statistics gives for all $q \neq 1$ a power law distribution, while the B-G exponential distribution is obtained only for $q$=1. This new theory has been successfully applied to a great number of nonextensive systems [7-9].

We should emphasize that not all nonextensive systems need Tsallis statistics to understand their behavior [1]. In the light of present understanding, it is unclear that which class of nonextensive systems requires Tsallis statistics for its statistical description, due to the physical meaning of the nonextensive parameter $q$ is still an open problem and the connection between the nonextensivity and the dynamics of a particle system, possibly leading to Tsallis rather than B-G statistics, is also unknown. A theorem proved by Almeida states recently that the canonical distribution function of the system is Tsallis distribution if and only if heat capacity of its environment (or the "heat bath") is assumed to be finite [10]. It is expressed by

$$1-q = -\frac{d}{dE}\left(\frac{1}{\beta}\right) \qquad (2)$$

with $\beta = 1/kT$ and $q$ a constant, where $E$ is the energy of the environment and $T$ the equilibrium temperature. This theorem gives $q \neq 1$ a physical interpretation about finite heat capacity of the "heat bath", leading to Tsallis statistics as a small "heat bath" statistics [11,12]. In this Letter, the nonextensivity in the non-isothermal plasma systems is analyzed on the basis of the generalized Maxwellian $q$-velocity distribution in the kinetic theory and the generalized Boltzmann equation in the frame of Tsallis statistics. We will give a mathematical expression of the nonextensive parameter $q$ based on the solid mathematical theory, which leads to a new physical explanation for $q \neq 1$ concerning the temperature gradient and the Coulombian long-range force on an electron. Our kinetic analyses are also appropriate to the description of the systems with Newtonian gravitational long-range interactions.

Systems of short-range interacting particles in thermal equilibrium are well known to have the Maxwellian velocity distributions. But the systems of particles with long-range interactions may be not so. It was reported experimentally that non-Maxwellian electron velocity distributions were measured in plasma that the Coulomb collisions are dominant, and turbulence has not developed significantly, where the plasma has been heated by intense microwaves so that very steep temperature gradient exists in it [13]. This is an inspired measurement because it indicates that the non-Maxwellian electron velocity distributions may be presented in the plasma where the Coulombian interactions play an important role. It has been showed recently that the standard Maxwell velocity distribution might provide only a very crude description for plasma physics, even in the limit of electronic collisionless, due to the long-range property of Coulombian interactions [14]. In fact, the Maxwellian velocity distribution function has been generalized in the nonextensive framework of Tsallis' statistics by the form [15]

$$f_q(\mathbf{v}) = nB_q \left(\frac{m}{2\pi kT}\right)^{\frac{3}{2}} \left[1 - (1-q)\frac{m\mathbf{v}^2}{2kT}\right]^{\frac{1}{1-q}} \tag{3}$$

where $B_q$ is a parameter only related to the nonextensive parameter $q$, $n$ is the number density of particles, $k$ is the Boltzmann's constant, $T$ is the temperature, $m$ and $\mathbf{v}$ is mass and velocity of the particle, respectively. When we take $q \to 1$, the standard Maxwellian velocity distribution is recovered from Eq.(3) correctly. In particular, Eq.(3) with $q<1$ has been strongly suggested by the results for electronic velocity distribution in a thermal plasma [14]. This generalized Maxwellian $q$-velocity distribution has been used to analyze the negative heat capacity [16] and Jeans instability of a self-gravitating system [17,18], the nonextensive transport property [19] and the distribution in conservative force field [20]. The general form of Eq.(3) of extending to include nonuniform systems with interparticle interactions was verified from the generalized Boltzmann equation and $q$-H theorem [21], which can be expressed by

$$f_q(\mathbf{r},\mathbf{v}) = nB_q \left(\frac{m}{2\pi kT}\right)^{\frac{3}{2}} \left[1 - (1-q)\frac{m(\mathbf{v}-\mathbf{v}_0)^2}{2kT}\right]^{\frac{1}{1-q}} \tag{4}$$

where $\mathbf{v}_0$ is the macroscopic entirety motion velocity, the density $n$ and temperature $T$ are now considered as space inhomogeneous functions. In this generalized Maxwellian $q$-velocity distribution, there is a thermal cutoff on the maximum value allowed for the velocity of an electron for $q<1$, $|\mathbf{v}-\mathbf{v}_0|_{max} = \sqrt{2kT/m(1-q)}$, whereas for $q>1$ without the thermal cutoff. It may be believed that the origin of the nonextensivity comes from the long-range interactions and the space inhomogeneity of the system. In the following context, the nonextensive parameter $q$ is regarded as a function of $\mathbf{r}$ so as to consider its contribution more sufficiently to the description for nonextensivity in the plasma system.

We consider the magnetic-field-free plasma in the $q$-nonextensive context. We use $f_q(\mathbf{r},\mathbf{v},t)$ to denote the $q$-velocity distribution function of electrons, while the distribution of ions can be considered invariable, then the behavior of the plasma is governed by the generalized Boltzmann equation [21]

$$\frac{\partial f_q}{\partial t} + \mathbf{v} \cdot \frac{\partial f_q}{\partial \mathbf{r}} + \frac{e}{m} \nabla \varphi \cdot \frac{\partial f_q}{\partial \mathbf{v}} = C_q(f_q) \tag{5}$$

where $C_q$, on the right hand side of this equation, is called the *q*-collision term, *e* is the electronic charge, *m* is its mass, $\varphi$ is the electric field potential, satisfying the Poisson equation

$$\nabla^2 \varphi = -4\pi e n \tag{6}$$

*Lima* et al has verified [21] that the solutions of the generalized Boltzmann equation (5) satisfy the generalized *q-H* theorem only if *q* >0 and evolves irreversibly towards the Tsallis' equilibrium distribution (the generalized Maxwellian *q*-velocity distribution [14 -17]) that, extending to nonuniform systems with interparticle interactions, is Eq.(4). When the *q-H* theorem is satisfied, the *q*-collision term $C_q$ vanishes and $f_q(\mathbf{r},\mathbf{v},t)$ evolves irreversibly towards the Tsallis' equilibrium distribution (4) and the generalized Boltzmann equation (5) becomes

$$\mathbf{v} \cdot \nabla f_q + \frac{e}{m} \nabla \varphi \cdot \nabla_v f_q = 0 \tag{7}$$

where we have used $\nabla = \partial/\partial \mathbf{r}, \nabla_v = \partial/\partial \mathbf{v}$. Or, Eq.(7) can be written in a more convenient form as

$$\mathbf{v} \cdot \nabla \ln f_q + \frac{e}{m} \nabla \varphi \cdot \nabla_v \ln f_q = 0 \tag{8}$$

We consider Eq.(4) in the form,

$$\ln f_q = \ln\left[ nB_q \left(\frac{m}{2\pi kT}\right)^{\frac{3}{2}} \right] - \sum_{i=1}^{\infty} \frac{1}{i}(1-q)^{i-1}\left[\frac{m(\mathbf{v}-\mathbf{v}_0)^2}{2kT}\right]^i \tag{9}$$

where the expression of power series is defined only if $-1 \leq (1-q)m(\mathbf{v}-\mathbf{v}_0)^2/2kT < 1$ is satisfied. This condition is equal to a thermal cutoff on the maximum value allowed for the velocity of an electron for *q*>1, $|\mathbf{v}-\mathbf{v}_0| \leq \sqrt{2kT/m(q-1)}$, and whereas for *q*>1,

$|\mathbf{v}-\mathbf{v}_0|<\sqrt{2kT/m(1-q)}$. Substituting Eq.(9) into Eq.(8), we have

$$\mathbf{v}\cdot\left\{\nabla\ln\left[nB_q\left(\frac{m}{2\pi kT}\right)^{\frac{3}{2}}\right]+\sum_{i=1}^{\infty}\mathbf{A}_{qi}(T,\nabla T)\left(\frac{m(\mathbf{v}-\mathbf{v}_0)^2}{2kT}\right)^i\right\}$$

$$-\nabla\varphi\cdot(\mathbf{v}-\mathbf{v}_0)\frac{e}{kT}\sum_{i=0}^{\infty}\left(\frac{(1-q)m}{2kT}(\mathbf{v}-\mathbf{v}_0)^2\right)^i=0 \tag{10}$$

where we have denoted

$$\mathbf{A}_{qi}(T,\nabla T)=(1-q)^{i-1}\frac{\nabla T}{T}+\left(\frac{1}{i}-1\right)(1-q)^{i-2}\nabla(1-q) \tag{11}$$

Substituting

$$\left[(\mathbf{v}-\mathbf{v}_0)^2\right]^i=\sum_{p+s+j=i}\frac{i!}{p!s!j!}(-2\mathbf{v}\cdot\mathbf{v}_0)^s v_0^{2j} v^{2p} \tag{12}$$

into Eq.(10), we have

$$\mathbf{v}\cdot\nabla\ln\left[nB_q\left(\frac{m}{2\pi kT}\right)^{\frac{3}{2}}\right]+\mathbf{v}\cdot\sum_{i=1}^{\infty}\mathbf{A}_{qi}\left(\frac{m}{2kT}\right)^i\times\sum_{p+s+j=i}\frac{i!}{p!s!j!}(-2\mathbf{v}\cdot\mathbf{v}_0)^s v_0^{2j} v^{2p}$$

$$-\frac{e\nabla\varphi}{kT}\cdot(\mathbf{v}-\mathbf{v}_0)\sum_{i=0}^{\infty}\left(\frac{(1-q)m}{2kT}\right)^i\times\sum_{p+s+j=i}\frac{i!}{p!s!j!}(-2\mathbf{v}\cdot\mathbf{v}_0)^s v_0^{2j} v^{2p}=0 \tag{13}$$

Because **r** and **v** are independent variables of each other, Eq.(13) is an identical equation for the velocity **v** and the sum of the coefficients of equal powers for **v** in Eq.(13) must be zero. We therefore find the equation of the zero-order terms of **v**

$$\frac{e\nabla\varphi\cdot\mathbf{v}_0}{kT}\sum_{i=0}^{\infty}\left[\frac{(1-q)m}{2kT}\right]^i v_0^{2i}=0$$

and then have

$$\nabla\varphi\cdot\mathbf{v}_0=0 \tag{14}$$

We will make use of Eq.(14) in the following calculations. The coefficient equation of the first-order terms of **v** is

$$\nabla \ln \left[ nB_q \left( \frac{m}{2\pi kT} \right)^{\frac{3}{2}} \right] - \frac{e\nabla\varphi}{kT} + \sum_{i=1}^{\infty} \left( \frac{m}{2kT} \right)^i v_0^{2i} \left[ \mathbf{A}_{qi} - (1-q)^i \frac{e\nabla\varphi}{kT} \right] = 0 \tag{15}$$

The equation of the second-order terms of **v** is

$$\mathbf{v} \cdot \sum_{i=1}^{\infty} \left( \frac{m}{2kT} \right)^i i(\mathbf{v} \cdot \mathbf{v}_0) v_0^{2(i-1)} \left[ \mathbf{A}_{qi} - (1-q)^i \frac{e\nabla\varphi}{kT} \right] = 0 \tag{16}$$

The equation of the third-order terms of **v** is

$$\mathbf{v} \cdot \sum_{i=1}^{\infty} \left( \frac{m}{2kT} \right)^i i\, v_0^{2(i-1)} \left[ v^2 + 2(i-1) \frac{(\mathbf{v} \cdot \mathbf{v}_0)^2}{v_0^2} \right] \left[ \mathbf{A}_{qi} - (1-q)^i \frac{e\nabla\varphi}{kT} \right] = 0 \tag{17}$$

The equation of the fourth-order terms of **v** is

$$\mathbf{v} \cdot \sum_{i=1}^{\infty} \left( \frac{m}{2kT} \right)^i i(i-1)(\mathbf{v} \cdot \mathbf{v}_0) v_0^{2(i-2)} \left[ v^2 + \frac{4(i-2)}{3!} \frac{(\mathbf{v} \cdot \mathbf{v}_0)^2}{v_0^2} \right] \left[ \mathbf{A}_{qi} - (1-q)^i \frac{e\nabla\varphi}{kT} \right] = 0 \tag{18}$$

The general form of the equation of the *l*th-order terms of **v**, *l*=2,3,4,5…, is given by

$$\mathbf{v} \cdot \sum_{i=1}^{\infty} \left( \frac{m}{2kT} \right)^i V_{l-1}(i, v_0, v^{l-1}) \left[ \mathbf{A}_{qi} - (1-q)^i \frac{e\nabla\varphi}{kT} \right] = 0 \tag{19}$$

where $V_l(i, v_0, v^l)$ denotes the sum of those terms containing $v^l$ in Eq.(12) by

$$V_l(i, v_0, v^l) = \sum_{\substack{p+s+j=i \\ (s+2p=l)}} \frac{i!}{p!\,s!\,j!} (-2\mathbf{v} \cdot \mathbf{v}_0)^s v_0^{2j} v^{2p} \tag{20}$$

The density distribution $n(\mathbf{r})$ of electrons can be determined by Eq.(15), while the nature of the nonextensive parameter $q$ is analyzed by Eq.(16), Eq.(17), Eq.(18)… or by Eq.(19). Our analyses undoubtedly follow the standard line in Boltzmann kinetic theory. We are interested mainly in the physics of $q$ and in the plasma system where the Coulombian interaction plays a dominant role, while turbulence has not developed significantly. In this case, the macroscopic motion velocity $v_0$ may be considered to be small. In the following analyses, we only consider its linear terms in the above equations (16-19), neglecting those terms that contain the powers of $v_0$ equal to and more than tow.

First, from Eq.(16) as well as Eq.(17), we have

$$\mathbf{A}_{q1} - (1-q)\frac{e\nabla\varphi}{kT} = 0 \tag{21}$$

and then we find

$$k\nabla T - (1-q)e\nabla\varphi = 0 \tag{22}$$

It is shown clearly in this equation that the nonextensive parameter is $q \neq 1$ if and only if the temperature gradient $\nabla T \neq 0$, which gives a clear physics of $q \neq 1$ concerning the nature of non-isothermal configurations of plasma systems with Coulombian long-range interactions. Furthermore, we can write

$$1 - q = \frac{k\nabla T \cdot d\mathbf{r}}{e\nabla\varphi \cdot d\mathbf{r}} \tag{23}$$

This gives a mathematical expression of the nonextensive parameter $q$.

Second, from Eq.(18) we have

$$\mathbf{A}_{q2} - (1-q)^2 \frac{e\nabla\varphi}{kT} = 0 \tag{24}$$

and then we find

$$\nabla q = 0 \tag{25}$$

It could be verified for all $l = 2,3,4\ldots$ in Eq.(19) that Eq.(23) and Eq.(25) are always obtained when we consider the linear approximation about $v_0$ only. If the macroscopic motion velocity $v_0$ is large enough, then turbulence will play a dominant role, which is beyond the scope of this paper.

Eq.(23) is a mathematical expression of the nonextensive parameter $q$, it gives a clearly physical meaning about temperature gradient and the Coulombian force on an electron in the non-isothermal plasma. If $\nabla T = 0$, the system becomes isothermal and we have $q=1$, which corresponds to the thermal equilibrium state for which B-G statistics has presented well description. While if $\nabla T \neq 0$, then $q \neq 1$, which corresponds to the case of Tsallis statistics. We therefore conclude that Tsallis statistics might deal with the non-isotherma nature in plasma systems with Coulombian long-range interactions. This result agrees with our previous conjectures on the physical meaning of $q$ in the discussions about the self-gravitating systems [18,22].

On the other hand, Eq.(23) can be written directly in the other form

$$1-q = \frac{kdT}{ed\varphi} \tag{26}$$

This is consistent with Eq.(2) obviously. Generally speaking, when we deal with a nonequilibrium system with a nonuniform temperature in it, its environment is space inhomogeneous. If the environment is considered space inhomogeneous, Eq.(2) is written immediately by

$$1-q = -k\frac{\nabla T \cdot d\mathbf{r}}{\nabla E \cdot d\mathbf{r}} \tag{27}$$

This is also consistent with Eq.(23). It means that we have provided one illustration for Almeida's theorem Eq.(2) from the kinetic theory of plasma systems, which indicates that Tsallis statistics might be a suitable statistics for the description of a nonequilibrium system with a temperature gradient in it.

We now write the generalized Maxwell $q$-distribution of electrons in the plasma system. As above, under the linear approximation of $v_0$, from Eq.(15) we have,

$$\nabla \ln\left[nB_q\left(\frac{m}{2\pi kT}\right)^{\frac{3}{2}}\right] - \frac{e\nabla\varphi}{kT} = 0 \tag{28}$$

and then, the density distribution is given by

$$n(\mathbf{r}) = n_0 \left(\frac{T(\mathbf{r})}{T_0}\right)^{\frac{3}{2}} \exp\left[\frac{e}{k}\left(\int \frac{\nabla\varphi(\mathbf{r})}{T(\mathbf{r})} \cdot d\mathbf{r} - \frac{\varphi_0}{T_0}\right)\right] \tag{29}$$

where we have let $B_{q(\mathbf{r})} = B_{q(\mathbf{r}=0)}$ due to Eq.(25), the integral constants $n_0$, $T_0$ and $\varphi_0$ denote the density, the temperature and the electric field potential at $\mathbf{r} = 0$, respectively. Substituting Eq.(29) into Eq.(4), we write the generalized Maxwell $q$-distribution of electrons as

$$f_q(\mathbf{r},\mathbf{v}) = n_0 B_q \left(\frac{m}{2\pi kT_0}\right)^{\frac{3}{2}} \exp\left[\frac{e}{k}\left(\int \frac{\nabla\varphi(\mathbf{r})}{T(\mathbf{r})} \cdot d\mathbf{r} - \frac{\varphi_0}{T_0}\right)\right] \left[1-(1-q)\frac{m(\mathbf{v}-\mathbf{v}_0)^2}{2kT(\mathbf{r})}\right]^{\frac{1}{1-q}} \tag{30}$$

If $\nabla T = 0$, we have $q = 1$ ($T(\mathbf{r}) = T_0$, $B_Q = 1$ as it should be), and the standard Maxwell distribution is recovered perfectly from Eq.(30).

In conclusion, we have studied the nonextensivity of a noneqilibrium thermal plasma system with Coulombian long-range interactions. We obtain for first time an analytic expression Eq.(23) of the nonextensive parameter $q$ based on the solid mathematical foundation about the generalized Boltzmann equation and the $q$-$H$ theorem and the generalized Maxwellian $q$-velocity distribution, which gives a clear physical explanation for $q \neq 1$ concerning the nature of non-isothermal configurations of plasma systems with Coulombian long-range interactions. We also obtain the result that is consistent with Eq.(2) and thus provide one illustration for Almeida's theorem from kinetic theory of the plasma system, which indicates that Tsallis statistics might be a suitable statistics for the description of a nonequilibrium system with a temperature gradient in it.

## Acknowledgments

I would like to thank Dr.W-S Dai for useful discussions. This work is supported by the project of "985" Program of TJU of China.